\documentclass[sigconf, review=false, anonymous=false]{acmart}

\usepackage{booktabs} 
\usepackage{xargs} 
\usepackage{xcolor}  
\usepackage[colorinlistoftodos,prependcaption,textsize=normalsize]{todonotes}

\usepackage{tabularx}

\usepackage{graphicx}

\usepackage{amsmath}
\usepackage{algorithmic}

\usepackage{array}

\usepackage{url}

\usepackage{comment}

\usepackage{import}

\newcommand{\code}{\texttt}

\copyrightyear{2018} 
\acmYear{2018} 
\setcopyright{licensedusgovmixed}
\acmConference[ICPE '18]{ACM/SPEC International Conference on Performance Engineering}{April 9--13, 2018}{Berlin, Germany}
\acmPrice{15.00}
\acmDOI{10.1145/3184407.3184421}
\acmISBN{978-1-4503-5095-2/18/04}

\begin{document}
\title{Pattern-based Modeling of Multiresilience Solutions for High-Performance Computing}

\author{Rizwan A. Ashraf, Saurabh Hukerikar, and Christian Engelmann}
\email{[ashrafra, hukerikarsr, engelmannc]@ornl.gov}
\affiliation{
  \department{Computer Science and Mathematics Division,}
  \institution{Oak Ridge National Laboratory,}
  \streetaddress{1 Bethel Valley Road,}
  \city{Oak Ridge}
  \state{Tennessee}
  \country{USA}
}

\renewcommand{\shortauthors}{R. A. Ashraf, S. Hukerikar, and C. Engelmann}
\renewcommand{\shorttitle}{Pattern-based Modeling of Multiresilience Solutions for HPC}

\thanks{This manuscript has been authored by UT-Battelle, LLC under Contract No. DE-AC05-00OR22725 with the U.S. Department of Energy. The United States Government retains and the publisher, by accepting the article for publication, acknowledges that the United States Government retains a non-exclusive, paid-up, irrevocable, worldwide license to publish or reproduce the published form of this manuscript, or allow others to do so, for United States Government purposes. The Department of Energy will provide public access to these results of federally sponsored research in accordance with the DOE Public Access Plan (http://energy.gov/downloads/doe-public-access-plan). \\}

\begin{abstract}
Resiliency is the ability of large-scale high-performance computing (HPC) applications to gracefully handle errors, and recover from failures. In this paper, we propose a pattern-based approach to constructing resilience solutions that handle multiple error modes.
Using resilience patterns, we evaluate the performance and reliability characteristics of detection, containment and mitigation techniques for transient errors that cause silent data corruptions and techniques for fail-stop errors that result in process failures. We demonstrate the design and implementation of the multiresilience solution based on patterns instantiated across multiple layers of the system stack. The patterns are integrated to work together to achieve resiliency to different error types in a performance-efficient manner.
\end{abstract}



\maketitle

\section{Introduction}
\label{sec:Introduction}

Resiliency solutions provide capabilities for high-performance computing (HPC) applications to deal with the effects of different types of errors, and recover from failures. Resiliency is becoming an increasingly important attribute for HPC systems and their applications, as systems of unprecedented scale and complexity are designed and deployed for running advanced scientific simulation, modeling, big-data analysis and machine learning applications. 
The continuous occurrence of faults is typical on the fastest supercomputing systems today due to reduction in reliability of individual system components caused by shrinking process technology, and operation at low voltage. As a consequence of these disturbing trends, and the growing complexity of the architectures and the software environment of HPC systems, future extreme-scale systems are projected to encounter frequent, persistent and erratic errors of different types \cite{DoD:ResilienceReport}. Therefore, the development of comprehensive resiliency solutions is critical to deliver sustained high performance for scientific applications.

Many of the resilience solutions in use today are designed to support a specific fault model. However, fault analyses indicate that modern HPC systems experience multiple types of error events with different levels of severity in terms of the application's ability to produce a correct solution and their impact on performance. Transient errors that cause silent data corruptions in the application state may result in outcomes ranging from loss in precision to wildly incorrect results. Unrecoverable errors often result in fail-stop behavior, which is fatal for the application program. Therefore, HPC applications require multiresilience solutions that provide comprehensive protection against multiple modes of errors. These solutions must be constructed systematically through integration of various techniques to detect and gracefully handle the error events without sacrificing application performance.

In this paper, we demonstrate a performance-oriented approach to the design and implementation of software-based multiresilience solutions. We leverage resilience design patterns, which we developed in previous work \cite{Hukerikar:2017}, to identify and evaluate techniques for detection, containment and mitigation for specific error modes. We explore an implementation of the complete multiresilience solution, in which patterns are instantiated across multiple layers of the system stack and work together to achieve required levels of performance and resiliency. The pattern-based approach also enables global optimization of the solution, avoiding costly over-protection and emphasizing end-to-end application performance. With this approach, we make the following significant contributions: 
\begin{itemize}
\item We demonstrate a novel approach based on resilience design patterns that systematically explores techniques with different performance and reliability characteristics, and enables the design of comprehensive multiresilience solutions through composition of patterns.
\item We design a cross-layer multiresilience solution from conception to implementation using patterns for linear solver methods. The solution is implemented by instantiating algorithmic patterns to work in concert with patterns incorporated in the message-passing layer of a parallel application.
\item We present a detailed experimental evaluation of our pattern-based solution that assesses the interdependencies between patterns for hard and soft errors, and characterizes the performance of the complete multiresilience solution. 
\end{itemize}     

\section{Background: Design Patterns for Resilience}
\label{sec:Background}

Design patterns describe a generalizable solution to a recurring problem that occurs within a well-defined context. Many of the HPC resilience solutions in use today are based on a fixed set of techniques that are repeatedly found in various solutions. We mined existing solutions, which have been used in HPC environments to confront faults, errors and failures, to discover patterns \cite{Hukerikar:2017}. 

Each resilience design pattern consists of a set of activation and response interfaces, and a behavior specification, which describes the semantics of how the pattern handles a fault event and its consequences. While the patterns are not finished designs that can be transformed into code, they outline the strategies for detecting a fault, error, or a failure, limiting its propagation, and mitigating its impact through recovery or masking. Therefore, discrete implementations of the same pattern may have different levels of performance and reliability characteristics.
We presented the resilience design patterns in a catalog \cite{RDP:Spec}, which organizes the patterns in a layered hierarchy. Based on the insight that HPC resilience has two important aspects, namely the forward progress of an application and the consistency and fidelity of an application's data, the catalog broadly categorizes patterns into state and behavioral patterns.
 
\subsection{State Patterns}
The state patterns describe the protection domain of a resilience solution. These patterns encapsulate the particular aspects of an application's state. The careful scoping of the protection domain enables defining the resilience behavior in a modular fashion for the specific domain captured by the state pattern. 
The selection of the state pattern also helps define the containment scope, i.e., the scope of how far a fault or error event propagates.   
The state patterns have been classified into: (1) \textit{Static State} pattern, which encapsulates the application data that is computed once in the initialization phase and is unchanged thereafter, (2) \textit{Dynamic State} pattern, which describes the changing application state as the application progresses, (3) \textit{Environment State} pattern, which includes the state necessary to perform the computation, i.e., program code, environment variables, libraries, etc, and (4) \textit{Stateless} pattern, which defines null state, enabling designers to create solutions that define behavior without predefined scope.

\subsection{Behavioral Patterns}
The behavioral patterns identify common detection, containment, mitigation techniques that enable an application or the system that instantiates and implements these patterns to cope with the presence of fault, error, or failure events. These patterns have been classified hierarchically describing solutions from abstract to concrete. The categories include:
(1) \textit{Strategy} patterns, which describe high-level solutions for fault treatment, error or failure recovery and compensation. 
(2) \textit{Architecture} patterns, which convey specific methods necessary for the construction of a resilience solution. These patterns are a sub-class to the strategy patterns.
Both the architecture and strategy patterns are organized by the types of event (fault, error or failure) they handle and the specific actions taken to handle an event.
(3) \textit{Structural} patterns, which provide concrete description of a solution that is intended to guide the implementation of the resilience solution. These patterns describe specific solutions for fault monitoring and prediction, the forward and backward checkpoint recovery, and patterns that describe the specific ways of applying the redundancy approaches.   

\section{Pattern-based Modeling of Multiresilience Solutions}
\label{sec:Modeling}

Among the difficult challenges that face HPC application developers and system designers is the emergence of different types of errors caused by intermittent or permanent faults, gradual degradation and system aging related effects. Na\"ively stacking multiple resilience solutions often leads to overprotection,  and is exorbitantly expensive in terms of the overhead to application performance, since each solution for each fault types imposes its own overhead. In order for applications to harness the capabilities of modern extreme-scale HPC systems, comprehensive resilience solutions must be designed and implemented thoughtfully and methodically.  

In this section, we explore the construction of multiresilience solutions for HPC applications using the resilience design patterns to guide the selection of techniques for detection, containment and mitigation, and for optimization of the protection domain and overall application performance. As a case study, we choose an iterative linear solver since such solver methods are at the heart of most scientific simulation and modeling applications. 
Our design of a multiresilience solution aims to address two of the most prevalent fault models in modern HPC systems, including transient errors that result in silent data corruptions (SDCs) and unrecoverable errors, which in the context of parallel applications based on communicating processes tends to be fatal causing failure of the entire application. The scope of this work is limited to faults which affect application correctness or completion, and does not include faults which explicitly affect performance, such as due to a slow parallel file system, a congested network, a code bug during application development, etc. 

The patterns in our catalog \cite{RDP:Spec} each have significantly different performance efficiency and complexity characteristics. Therefore, we quantitatively evaluate their viability for our context. The complete design of the multiresilience solution entails combining the selected patterns and optimizing their interactions to ensure complete protection to both error types while delivering greater end-to-end application performance. 

\subsection{Patterns for Soft Error (SE) Detection and Mitigation}
Soft errors that cause SDCs affect the convergence properties of a linear solver method. A typical linear solver method solves a system of equations of the form $Ax=b$, in which the matrix $A$ and right hand side vector $b$ are known, and we solve to determine vector $x$. When affected by SDCs during the solution phase, the corruption may cause unbounded numerical errors in the outcome, slower convergence to solution, or the premature termination of the solver. However, application specific knowledge about the data structures and the algorithm used is required to detect SDCs in the solver and mitigate their effects. 

\textbf{\textit{State Patterns for SE resilience:}}  In the linear solver, the operand matrix $A$ and right hand side vector $b$ are initialized during the setup and remain unchanged throughout the computation. Therefore, variables $A$ and $b$ are encapsulated in \emph{static} state patterns. The solution vector $x$ is updated as the solver makes progress, until the linear solver converges. The remainder of the variables, which includes matrix index structures, pointers, loop counters, etc, are included in an \emph{environment} state pattern. The separation of the linear solver state into these distinct state patterns enables the exploration of different behavioral patterns for detection and recovery that leverage the properties of each state pattern. 

\textbf{\textit{SE Detection Patterns:}} For detecting the presence of SDCs in application state encapsulated by the \emph{static} state patterns, the behavioral patterns may take advantage of the invariance property of this state from initialization until the solver converges. Between the behavioral strategy patterns that offer \emph{recovery} and \emph{compensation}, the latter is more suitable based on the insight that the protection domain is contained in a static pattern, and therefore the redundant information required for detection is also computed at initialization. In contrast, the \emph{recovery} behavioral pattern incurs the overhead of preserving a checkpoint of its protection domain at periodic intervals, which in the case of a static state pattern is unnecessary. 

For the solver's protection domain contained in the \emph{dynamic} state pattern, the detection of SDCs requires insight into the changing nature of the dynamic state, which is a capability supported by the \emph{diagnosis} pattern, and specifically the \emph{monitoring} pattern. This pattern uses cause-effect analysis to infer the presence of faults or errors. For the linear solver application employing an iterative algorithm, the residual error in the solution indicates how close the solver is from a correct solution. 

An alternative pattern for detection of SDC that we considered is the \emph{compensation} pattern, which is realized as a \emph{n-modular} pattern. However, SDC detection using this pattern requires at least 2x computation and consumes additional communication bandwidth at large-scale resulting in high performance overheads. 

\textbf{\textit{SE Mitigation Patterns:}} The patterns for mitigating the impact of SDCs on the linear solver application  seek to ensure that the solver converges to a correct solution despite SDCs. For the protection domain scope in the \emph{static} state pattern, we apply a \emph{compensation} strategy pattern, which is realized using a \emph{redundancy} architectural pattern. Rather than implement this as a \emph{n-modular redundancy} pattern, we select the \emph{forward error correction} pattern to leverage the structure of the matrix and vector variables. 

For the variable state in the \emph{dynamic} state pattern, i.e., the solution vector, we must select the \emph{recovery} pattern; any \emph{compensation} strategy pattern is not viable due to the dynamic nature of the protected state and the exorbitant cost of instantiating modular redundancy pattern. The rollback recovery pattern derivative structural pattern is suitable for this context. This solution entails periodically preserving the \emph{dynamic} state pattern to persistent storage, which incurs an overhead to the application. However, by limiting the scope of the protection domain to the \emph{dynamic} state pattern, the \emph{recovery} pattern is applied to the solver computation only. 

\subsection{Patterns for Hard Error (HE) Detection and Recovery}
Parallel implementation of linear solver uses distributed memory model with message passing to distribute the problem over multiple processes that run on a number of compute nodes of a HPC system. In the utilized message passing interface (MPI), a communicator is a logical collection of processes that can send messages to each other. The processes use point-to-point or collective primitives to distribute data, exchange partial results and synchronize. The occurrence of an unrecoverable error causes MPI calls made by any process in the communicator to block indefinitely. Most MPI library implementations cannot stabilize after the failure of any one process in the communicator,  causing the remaining processes to deadlock, which renders the parallel application incapable of forward progress. 

\textbf{\textit{State Patterns for HE resilience:}}
To enable an MPI-based application to survive the occurrence of a hard error, the MPI library implementation must guarantee that it will stabilize the communicator itself following the process failure caused by the hard error, and the application must resolve the loss of partial application state that was resident on the failed process.
Therefore, in defining the scope of the protection domain, we encapsulate the MPI library and its runtime into an \emph{environment} state pattern.  
Much of the variable state associated with linear solver is distributed by partitioning the data in a block manner, i.e., row-wise chunks of matrices and vectors are distributed to processes. 
When a process failure occurs, all the variable state associated with process is lost. Accordingly, we encapsulate the state of each process in the MPI communicator into \emph{dynamic} state patterns. In this case, there is no advantage to encapsulating the individual variables into different state patterns. 
Based on such scoping of the MPI-based application's global state, the behavioral patterns for resolving the deadlock in the MPI communicator following a process failure may be applied separately from the behavioral patterns that resolve the loss of part of the solver's operands, i.e., the solution matrix and vector state.  

\textbf{\textit{HE Detection Patterns:}}
For the protection domain scope contained by the \emph{environment} state pattern, i.e., the MPI communicator, the detection pattern must determine which processes within a communicator have failed. The \emph{dynamic} state does not require an explicit detection pattern, since any unrecoverable error in the process state propagates to the environment state pattern.  

The failure detection pattern for the the \emph{environment} state pattern must be robust, and may take a proactive or a reactive approach. Typical detection techniques use periodic signaling to detect the liveliness of neighboring processes, or by building consensus between the processes alive. For scalability, failure detection may also be local, where failure detection is initiated only among the neighbors of a process.
To accomplish detection, we select a \emph{consensus} structural pattern, which is a derivative of the \emph{decentralized detection} strategy pattern. For its implementation, if an application requires timely notification of failure to all processes, then a proactive approach may involve strategic placement of collective operations inside the code, such that a failure is detected early on and costly re-computation is avoided. However, this can lead to high synchronization overheads, especially if there are no failures. Another approach is to wait for error notification by a collective operation in the code, i.e., a reactive approach. 

The implementation of the pattern is realizable using the failure detection primitives offered by the User-Level Failure Mitigation (ULFM)~\cite{ULFM:IJHPCA} implementation of an extended MPI. ULFM provides primitives such as \code{MPI\_COMM\_AGREE}, \code{MPI\_COMM\_REVOKE}, \code{MPI\_COMM\_FAILURE\_ACK}, and \code{MPI\_COMM\_FAILURE\_GET\_ACKED} to facilitate detection. Whereas, notification of failures to other processes is propagated through constructs such as \code{MPI\_ERR\_REVOKED} and \code{MPI\_ERR\_PROC\_FAILED}. It should be pointed that use of ULFM constructs require that MPI application changes the default error handler \code{MPI\_ERRORS\_ARE\_FATAL}. 

\textbf{\textit{HE Recovery Patterns:}}
The mitigation of a parallel application must be concerned with the \emph{environment} and \emph{dynamic} state patterns. The mitigation of the \emph{environment} entails stabilizing the MPI communicator following the detection of a process failure. While the \emph{dynamic} state pattern doesn't explicitly use a pattern for detection, its mitigation is critical since it encapsulates part of the parallel application's state, which is lost upon occurrence of a failure. 

To recreate the failed MPI communicator, the \emph{recovery} strategy pattern instantiated as a \emph{reconfiguration} architectural pattern is suitable. This pattern may be instantiated using the \emph{rejuvenation} or \emph{reinitialization} structural pattern.
An implementation of this pattern using the ULFM extensions to MPI would use \code{MPI\_COMM\_SHRINK} primitive to isolate a failed process from the MPI communicator used by the application. The key benefit of applying this pattern is that it offers the opportunity to resume the application without the need to resubmit the job to the scheduler in an HPC system.

The hard error mitigation for the \emph{environment} state pattern may also be accomplished by applying the \emph{compensation} strategy pattern. The instantiation of this pattern substitutes the failed process with another from a pool of spare processes. The use of this pattern as opposed to rejuvenation pattern eliminates the need to redistribute the workload among surviving processes which can be time-consuming and strongly application dependent. Besides, some HPC applications have strict requirements on the number of processes due to problem decomposition restrictions or memory pressure on nodes restricting additional workload.

The compensation pattern can be implemented by spawning processes at runtime called cold spares or allocating spare processes during initialization called hot or warm spares depending on how they are used.
Hot spares perform active concurrent execution, whereas warm spares are initialized and do nothing until a failure takes place. Warm spares do not maintain any dynamic or static state until they are put into service. There is also no need to have a spare for every process, if only a few failures are expected during the execution of the application. On the other hand, a hot spare is required for every process in the application to sustain arbitrary process failures.  This is analogous to redundant computation and does not require the checkpoint pattern for state recovery. However, processes lost over time result in loss of redundancy. Thus, use of hot spares can lead to significantly higher overheads and the in-ability to sustain failures of lost redundant processes as compared to an approach based on warm or cold spares. In general, spares are useful for applications which perform compute-intensive data distribution plan at the start of the application.

\begin{figure} [tp]
\centering
\includegraphics[width=\linewidth]{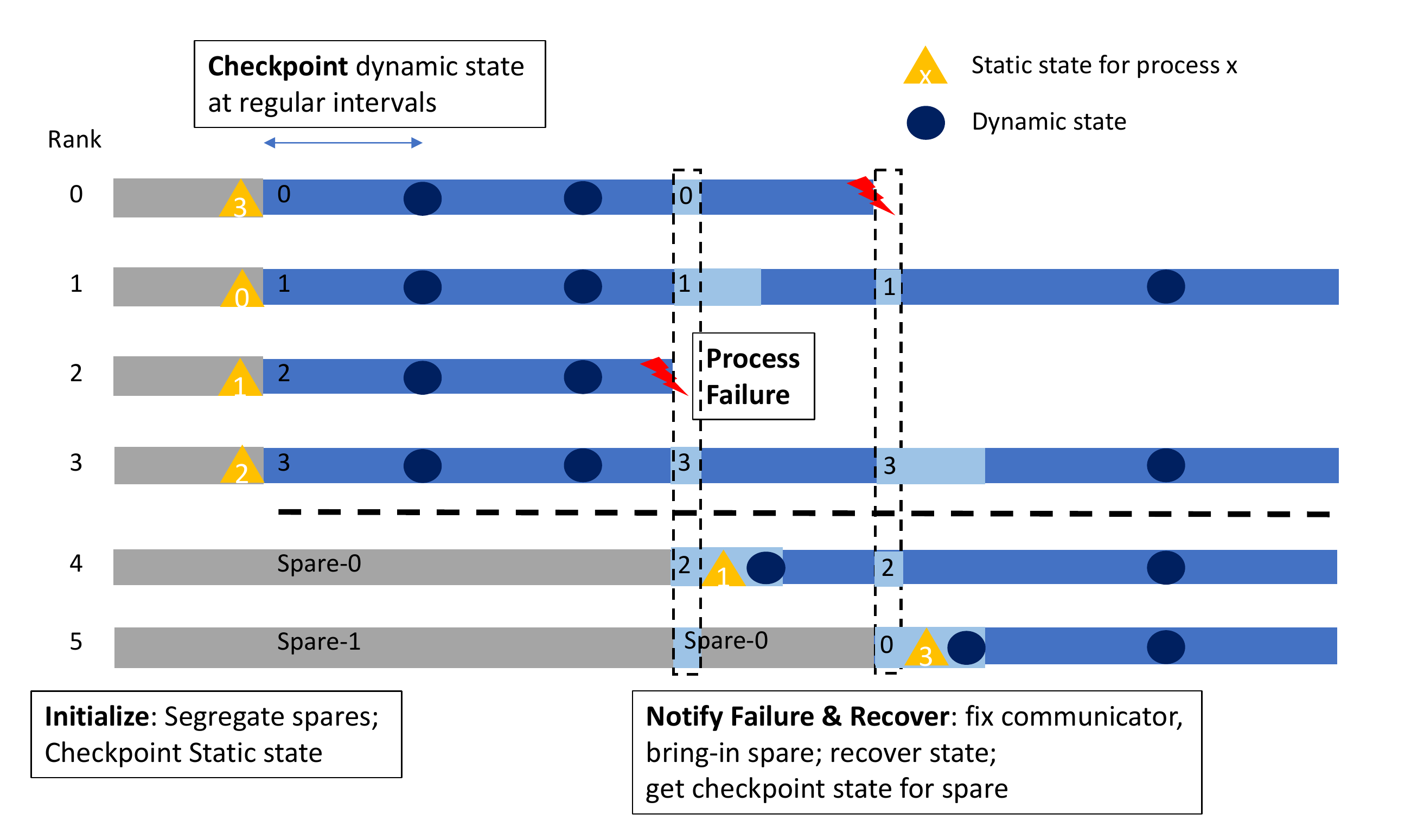}
\caption{Checkpoint restart pattern using warm spares.}
\label{Fig:PF-Resilience}
\end{figure}

When applying either the rejuvenation or compensation pattern for mitigation of the \emph{environment} state pattern, the \emph{dynamic} state pattern of the failed process must also be mitigated. Since the complete process state is lost in the event of a failure, a \emph{recovery} pattern is appropriate, specifically the \emph{checkpoint restart} pattern. 
We use in-memory checkpointing on a neighbor process to implement the checkpoint restart pattern, which efficiently leverages point-to-point communication between compute nodes rather than commit the checkpoint to a parallel file system~\cite{SCR:SC10}. The in-memory checkpointing is feasible because we have carefully defined the scope of the state patterns thereby limiting the memory overhead required for the checkpoints.
The use of \emph{checkpoint restart} pattern with warm spares strategy to mitigate process failures is illustrated in Figure~\ref{Fig:PF-Resilience}.

\section{Implementation of Pattern-based Multiresilience for Linear Solver}

\begin{footnotesize}
\begin{table*}
  \caption{Design Summary for Multiresilient FT-GMRES solver.}
  \label{tab:design}
  \begin{tabularx}{\linewidth}{m{0.08\linewidth}|m{0.07\linewidth}|m{0.29\linewidth}|m{0.18\linewidth}|m{0.28\linewidth}}
    \toprule
    \textbf{Fault Model} & \textbf{Pattern Class} & \textbf{Choices} & \textbf{Selection} & \textbf{Reason} \\
    \hline
    \emph{Soft Error}  & State  & Dynamic, static, environment & Dynamic & Static state corruption is detectable, environment corruption transforms to process failure \\ \cline{2-5}
      & Detection & Monotonicity, bounded compute, checksums & Bounded compute & About 14 times lower overhead compared to monotonicity pattern \\ \cline{2-5}
      & Recovery & \textit{Solver state:} restart inner, abort inner \& restart outer &  Restart inner & Low detection latency \\ \cline{3-5}
      &  & \textit{Variable state:} checkpoint, checksums & Checkpoint & Less computational overhead \\ \hline

    \emph{Hard Error} & State & Dynamic, static, environment & Dynamic, static, environment & Process failures are fatal for application \\ \cline{2-5}
     & Detection  &  Proactive, reactive & Proactive & Low overhead \& collective in every iteration \\ \cline{2-5}
     & Recovery   &  \textit{Solver state:} restart inner, abort inner \& restart outer  & Abort inner \& restart outer & Checkpoint state of outer \\ \cline{3-5}
     &  & \textit{MPI environment:} cold/warm spares, rejuvenate & Warm spares & Workload distribution \\ \cline{3-5}
     &  & \textit{Variable state:} checkpoint-restart, diskless checkpoint, interpolation & Checkpoint-restart & Less computational overhead \\
   \bottomrule
\end{tabularx}
\end{table*}
\end{footnotesize}

To demonstrate the design of multiresilience solution for a linear solver application encompassing soft and hard error resilience patterns, we consider the Generalized minimal residual (GMRES) method as a case study. 
We enumerate the design patterns for defining the protection domain, detection and mitigation patterns for each error class, and subsequently justify our selections. The suitability of patterns working in combination to provide a holistic multiresilience solution is an important consideration.

The GMRES method, which underlies many scientific computing applications, is a Krylov subspace method for the iterative solution of large sparse non-symmetric linear systems \cite{Saad:1986}. 
The \textit{flexible} GMRES extends this solver to permit the pre-conditioner to be changed every iteration \cite{Saad:1993} using inner-outer iterations; the ``inner" solve step preconditions the ``outer" flexible iteration. 
The inner-outer solver structure was leveraged to design Fault Tolerant GMRES (FT-GMRES) \cite{Hoemmen:2011} which provides robustness in the presence of unbounded errors. The FT-GMRES algorithm is designed to reach eventual convergence in the presence of soft errors, i.e., the solver produces a correct outcome at the cost of needing additional iterations to arrive at the right answer. It divides the computation into reliable and unreliable phases, i.e., \textit{selective reliability}. There is no assumption of reliability in the inner solver, i.e., it may return incorrect results as long as the solution is completed in finite time. 
On the other hand, the outer solver needs to be reliable, which is feasible since most of the time is spent in the inner solver.
The outer solver can also detect invalid values within the solution vector and replace them with arbitrary values for forward progress of the solver. The flexible inner-outer iterations have the property that the dimension of the Krylov subspace grows at each outer iteration, which guarantees eventual convergence.
 
\textbf{Implementation of SDC Detection and Recovery Patterns} 
We resolved to applying the monitoring structural pattern for detection of SDCs. The pattern implementation entails defining bounds on values produced during critical computational phases of GMRES. Specifically, an orthogonalization process based on the Arnoldi method is utilized to find orthonormal basis of the Krylov subspace~\cite{Saad:1986}. The basis is used to approximate the solution at each iteration of the solver. 
The projections produced during this critical computational phase are bounded by the upper bound of Frobenius norm of the input matrix. In a parallel implementation, the comparison of the projection length with the Frobenius norm is performed locally. This pattern implementation is referred to as the \emph{bounded compute} pattern.
An alternate algorithmic implementation of this pattern is possible using a monotonicity violation check. This implementation uses sparse matrix-vector multiplication (SpMVM) for calculating residue. While more generally applicable, it incurs as much as 14 times higher overhead in comparison to the implementation based on bounded compute pattern.

For implementing the recovery pattern for SDCs, we use an instantiation of the \emph{rollback} pattern. When applying this pattern, we leverage the inner-outer solver structure that expects high reliability from only the outer solver phases by scoping the rollback recovery pattern to exclude the inner solver phase. Our implementation creates local in-memory checkpoints.

\textbf{Implementation of HE Detection and Mitigation Patterns} 
Hard errors in an MPI-based implementation of the GMRES solver causes failure of the affected MPI process. For implementing the process failure detection and mitigation patterns, we embed ULFM primitives in the GMRES solver implementation. 

For detection of process failure within the MPI communicator, we leverage the collective SpMVM operations performed in every iteration of the GMRES solver. Detection is done using the returned error codes from MPI collective operations, which are caught by the error handler for the MPI communicator. Since no additional application code is required, the overhead of the detection pattern's implementation is negligible. For the mitigation of process failures, we implement a compensation strategy pattern, which instantiates a redundancy pattern. The implementation entails the creation of a pool of spare processes, which replace the failed ranks in the communicator; this implementation avoids the need to redistribute workload. 
For the recovery of the dynamic application state after a failure, we consider an algorithm-based compensation pattern that uses linear interpolation of known correct values to mitigate the affected state. While this method has no overhead during error-free operation, it causes slower convergence of the solver. Therefore, in our solution we apply the rollback recovery pattern for the static and dynamic state recovery after process failures. Its implementation creates in-memory checkpoints  on neighboring processes. 

\textbf{Implementing Multiresilience Solution for FT-GMRES}
Table~\ref{tab:design} summarizes our design choices for supporting multiresilience in FT-GMRES solver, describing the various resilience patterns considered for each fault model, the patterns selected for our multiresilience solution as well as the justification for their selection. 
Based on our empirical evaluation, we select algorithm-based instantiations of the patterns for SDC detection and recovery since they incur low performance overheads. However, the coverage of these pattern instances risks the possibility of prolonged execution on account of additional solver iterations.
The selection of rollback pattern for state recovery while handling process failures unintentionally causes the SDC detection pattern to be invoked more often than intended by the application programmer during re-execution after rollback. However, interaction of the rollback pattern with the SDC detection pattern limits propagation of SDC by preventing incorrect state from being captured during checkpointing.

\section{Experimental Setup and Evaluation}
\label{sec:ExperimentalEval}

The FT-GMRES is implemented using the Trilinos 12.6.4 framework~\cite{Trilinos} and uses the Tpetra package for parallel linear algebra operations such as SpMVM, vector dot products, etc.
We use ULFM release 1.1, which is derived from Open MPI-1.7.1, for process failure detection, notification, and rebuilding failed communicators.

\textit{Test problem and configuration:}
We solve a linear problem generated by discretizing a regular 3-D mesh using the Intrepid package~\footnote{https://trilinos.org/packages/intrepid/} in Trilinos framework.
The generated sparse matrix has 6,967,871 rows and 186,169,411 non-zero elements. We use the same problem size while scaling the number of processes, which causes the size of per-process checkpoint to shrink with increasing process counts. 
We fix the number of iterations of the inner solver of the FT-GMRES to 25, and number of iterations of the outer solver to 20 for a maximum iteration count of 500. 

\textit{Evaluation platform:}
We use a  Linux cluster with 40 compute nodes interconnected with a dual-bonded 1 Gbps Ethernet.
Each compute node has two AMD Opteron processors (a total of 24 cores) and 64 GB memory, for a total of 960 processor cores.
The switches support fully non-blocking point-to-point bandwidth of 215 MB/s.
We perform our fault injection experiments with 32, 64, 128, 256, and 512 processes, which are distributed across nodes of the cluster.  
The spare processes created by hard error recovery pattern instance are mapped to the last physical node. We maintain the same process mapping for all experiments to prevent application performance variability due to mapping. 

\textit{Soft error injection:}
The errors are injected at fixed intervals after every 10, 20, or 30 SpMVM operations. We randomly corrupt data elements produced after the completion of a SpMVM operation. 
These fault rates are chosen to understand the interaction with process failure resilience patterns; injecting a soft error after every 10 SpMVM operations means that the checkpointed dynamic state is affected more often than the injection after every 30 SpMVMs.

\textit{Process failure injection:}
To simulate hard errors and for reproducibility of results, the rank positions of MPI processes to be terminated are pre-selected. We also guarantee that the failed processes are on different physical nodes than the ones on which spare processes are mapped, and that sufficient spares are always available.  
Following these constraints, the processes are terminated randomly based on an exponential distribution with an average failure rate corresponding to time to complete at least 75 iterations. Under this assumption and based on Young's formulation, the checkpoint of dynamic state is performed at every iteration of the outer solver or after 25 iterations.
While other analytical models for estimating the checkpoint interval are available based on failure distributions~\cite{checkpoint-models:ICPE-2017}, we found our assumptions to be adequate for this work based on observed failure trends in current HPC systems. 

\section{Results}
\label{sec:Results}
\begin{figure} [t]
\centering
\includegraphics[width=\linewidth]{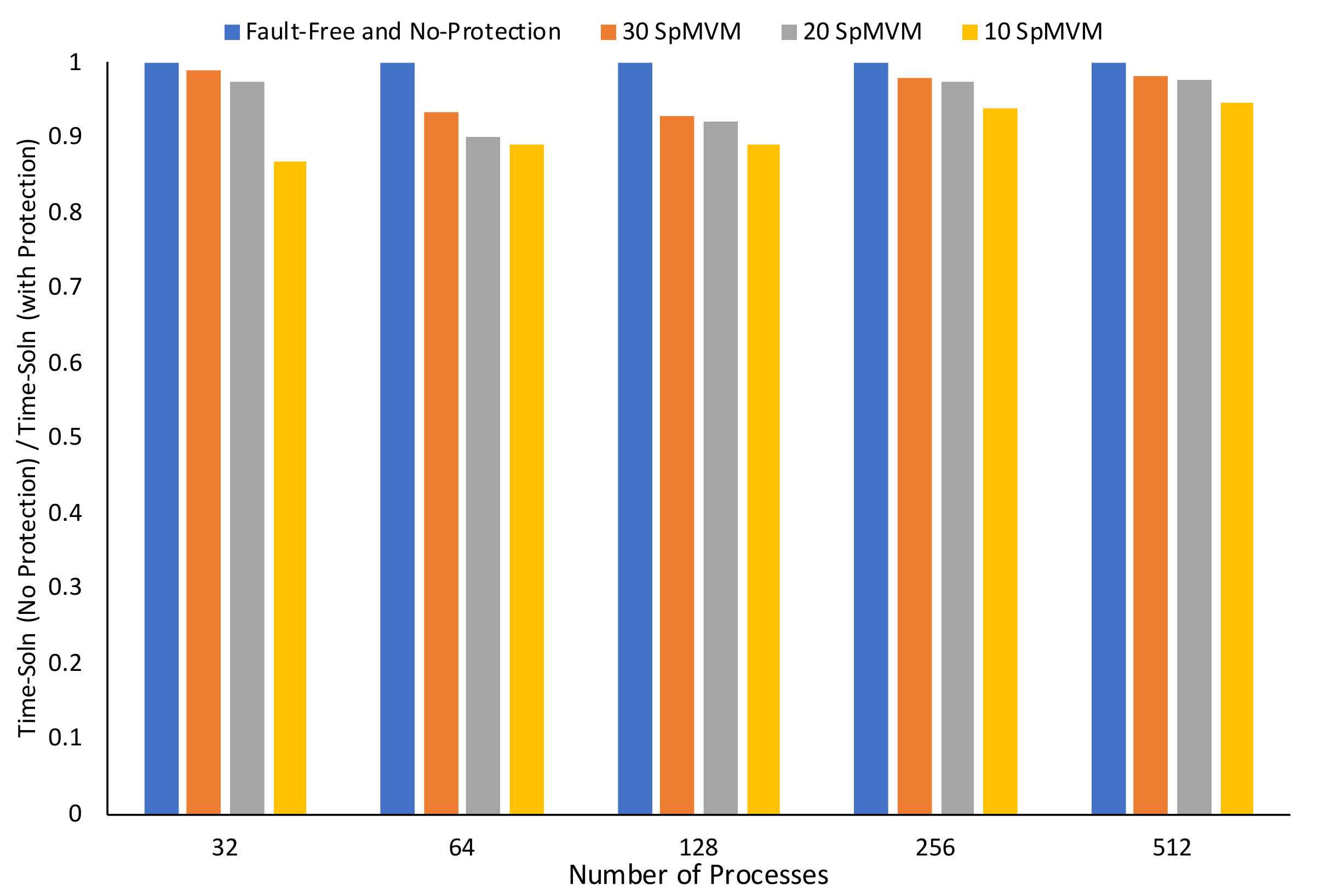}
\caption{Overheads of SE resilience with error injection after every 30 SpMVM, 20 SpMVM, and 10 SpMVM operations.}
\label{Fig:SEonlyOverall}
\end{figure}
\begin{figure} [t]
\centering
\includegraphics[width=\linewidth]{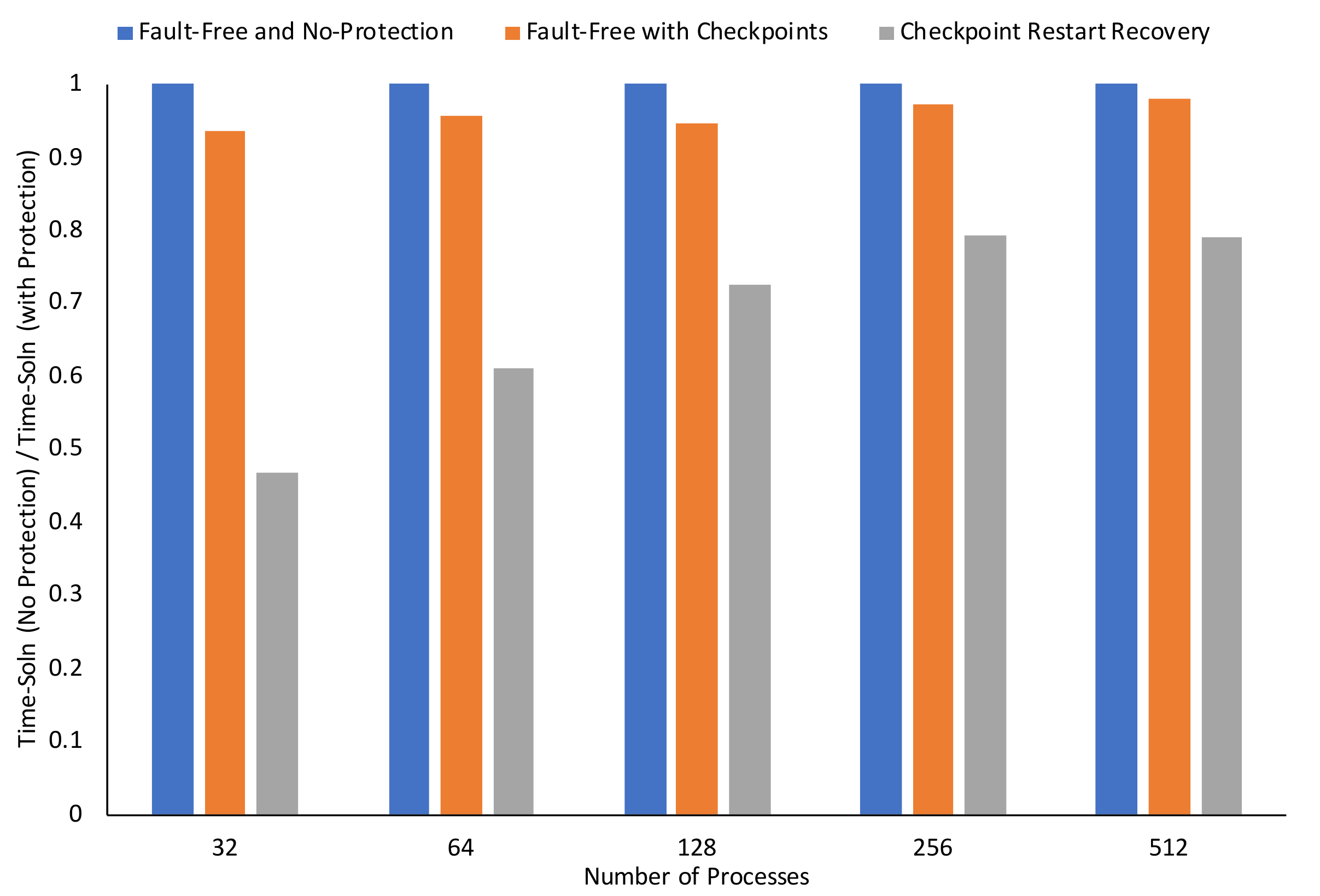}
\caption{Overheads of PF resilience with up to four process failures using checkpoint restart pattern and warm spares.}
\label{Fig:HEonlyOverall}
\end{figure}

The experiments are designed to analyze the reliability  and performance characteristics of individual patterns. We therefore evaluate the performance overheads of the patterns with different error rates and process counts. In each case, enough experiments are performed such that standard deviation is low. For instance, the coefficient of variation for all cases range between 0.01 and 0.15. 

The main motivation for these experiments is to assess the accuracy with which we can estimate the performance overhead for multiresilience solution from stand-alone soft error and hard error resilience experiments. This provides feasibility and constraints of various combination of patterns by analyzing their performance and resilience impact on the application.

\subsection{Soft Error (SE) Resilience}

The performance impact with three different soft error injection rates is quantified in Figure~\ref{Fig:SEonlyOverall}.
The y-axis shows the overhead on time-to-solution of providing resilience to errors, which includes the overheads of including detection and recovery patterns. 
In all cases, solver converged to a correct solution in the allotted time.

The soft error rates are chosen such that every inner solve operation is corrupted multiple times (10 SpMVM) or once (20 SpMVM), or after every other inner solve operation (30 SpMVM). 
Results indicate that high SDC rate leads to higher overheads across all processor counts.
The prime factor is the slower convergence of the solver in presence of more errors. 
The breakdown of overheads with two extreme error rates is listed in Table~\ref{tab:SE-overheads}.
The average number of extra iterations, represented as $N_{extra}$, consumed with error rate of 10 SpMVM is always higher than at 30 SpMVM.  
Mostly, the overhead due to extra computation is higher than the combined overheads of detection and recovery, which are represented as $t_{SDC-d}$ and $t_{SDC-r}$, respectively.
On the other hand, the SDC detection and recovery overheads tend to decrease with scale.

The results indicate a tradeoff between SDC detection overhead and the extra computation overhead. The use of monotonicity violation for SDC detection causes high overhead, but the number of extra iterations needed for convergence tends to decrease in comparison to detection based on the bounded computation pattern.
Overall, the utilized combination of SDC detection and recovery patterns results in minimum time to solution for FT-GMRES solver while providing resilience to soft errors.

\begin{footnotesize}
\begin{table}
  \caption{Breakdown of overheads related to SE resilience as a percentage of total time to solution.}
  \label{tab:SE-overheads}

  \begin{tabular}{|c|c|c|c|c|}
    \hline
    Processes  & \multicolumn{2}{c|}{$t_{SDC-d}+t_{SDC-r}$} & \multicolumn{2}{c|}{$N_{extra}$ [max] } \\
    \hline
                   &  \textit{30 SpMVM}  & \textit{10 SpMVM}                              & \textit{30 SpMVM} & \textit{10 SpMVM} \\ \hline

    $32$             &   $2.06\%$   & $8.91\%$ &  $30.1$ [$75$] & $35.5$ [$150$]  \\ \hline
    $64$             &   $6.45\%$   & $4.75\%$ &  $26.9$ [$50$] & $33.7$ [$75$]  \\ \hline
    $128$            &   $8.10\%$  & $8.75\%$  &   $25$  [$25$] &  $32.5$ [$125$]  \\ \hline
    $256$            &   $1.07\%$   & $1.94\%$ &   $32$  [$50$] &  $35.6$ [$125$]    \\ \hline
    $512$            &   $0.69\%$   & $0.72\%$ &  $28.3$ [$50$] & $34.6$ [$100$]  \\ \hline
  \end{tabular}
\end{table}
\end{footnotesize}

\subsection{Hard Error (HE) Resilience}

The performance impact of process failure resilience through checkpoint restart pattern and spare processes is quantified in Figure~\ref{Fig:HEonlyOverall}.
The overheads of checkpointing application state are indicative from `Fault-Free with Checkpoints' bar in Figure~\ref{Fig:HEonlyOverall}, where no process failures are injected.
This includes the overheads to perform initial checkpoint of static state and multiple checkpoints of dynamic state. 
These overheads tend to decrease with increasing number of processes, since the problem size is kept constant.
On average, the overheads range between 6.28\% to 1.98\%.

\begin{footnotesize}
\begin{table}
  \caption{Breakdown of overheads related to PF resilience as a percentage of total time to solution.}
  \label{tab:CR-overheads}
  \begin{tabular}{|c|c|c|c|c|}
    \hline
    Processes  & $t_{PF-x}$ & $t_{PF-r}$ & $t_{check}$ [\% dynamic] & $t_{recompute}$ \\
    \hline
    $32$       &   $0.02\%$          &     $17.1\%$      &    $28.1\%$ [$25.6\%$]            &  $10.9\%$   \\ \hline
    $64$       &   $0.03\%$          &     $9.4\%$       &    $18.5\%$ [$22.9\%$]            &  $13.4\%$   \\ \hline
    $128$      &   $0.04\%$          &     $5.4\%$       &    $12.9\%$ [$14.7\%$]            &  $12.9\%$   \\ \hline
    $256$      &   $0.02\%$          &     $1.9\%$       &    $7.5\%$  [$16.7\%$]            &  $13.5\%$   \\ \hline
    $512$       &   $0.05\%$          &     $1.2\%$       &    $5.1\%$  [$12.2\%$]            &  $16.2\%$   \\ \hline
\end{tabular}
\end{table}
\end{footnotesize}

The overheads of providing mitigation to process failures are indicated from `Checkpoint Restart Recovery' bar in Figure~\ref{Fig:HEonlyOverall}, where up to four independent process failures are injected based on the selected failure rate.
Significant overheads are notable at lower processor counts, while overheads tend to decrease at higher processor counts. 
These overheads include the following components: re-computation time, $t_{recompute}$, time to recover dynamic and static states using checkpoints, $t_{PF-r}$, time to fix MPI environment and include spares, $t_{PF-x}$, and the time to perform checkpoint of dynamic and static states, $t_{check}$.

Our analysis indicates that time to fix the distributed environment tends to be negligible, e.g., it varies between 0.02\% to 0.05\% of the total time to solution at our scale of experiments.
The dominant overheads are due to the following: $t_{PF-r}$, $t_{check}$ and $t_{recompute}$.
These overheads tend to be additive with the number of failures, whereas $t_{check}$ can be controlled via selection of checkpoint interval. 
The average values of these parameters are listed in Table~\ref{tab:CR-overheads}.
The checkpoint related overheads tend to decrease with scale from as high as 45\% at 32 processes to as low as 6\% at 512 processes, since the workload is kept constant.
The recovery overhead is mostly consumed by the time to deliver application state to the spare process which is put in service, i.e., communication of the spare with the neighbor of the failed process (see Figure~\ref{Fig:PF-Resilience}). 
Further investigation shows that significantly more time is spent in restoring static state as compared to dynamic state, which emphasizes the need to carefully scope the static and dynamic state patterns. 

\begin{figure} [tp]
\centering
\includegraphics[width=\linewidth]{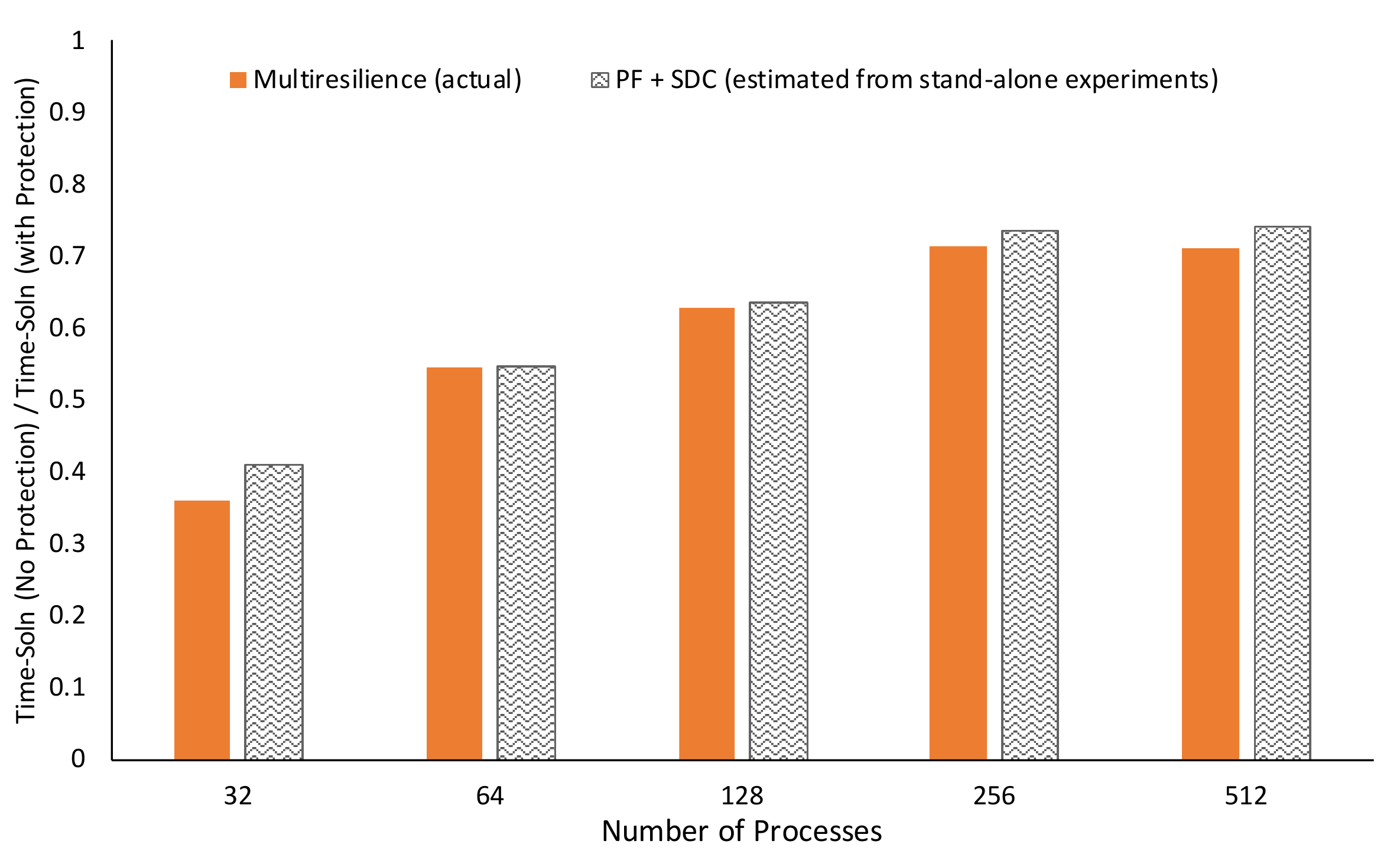}
\caption{Comparison of normalized times in case of multiresilience solution with estimated times from stand-alone PF resilience and SE resilience experiments.}
\label{Fig:HESEOverall}
\end{figure}

\begin{figure} [tp]
\centering
\includegraphics[width=\linewidth]{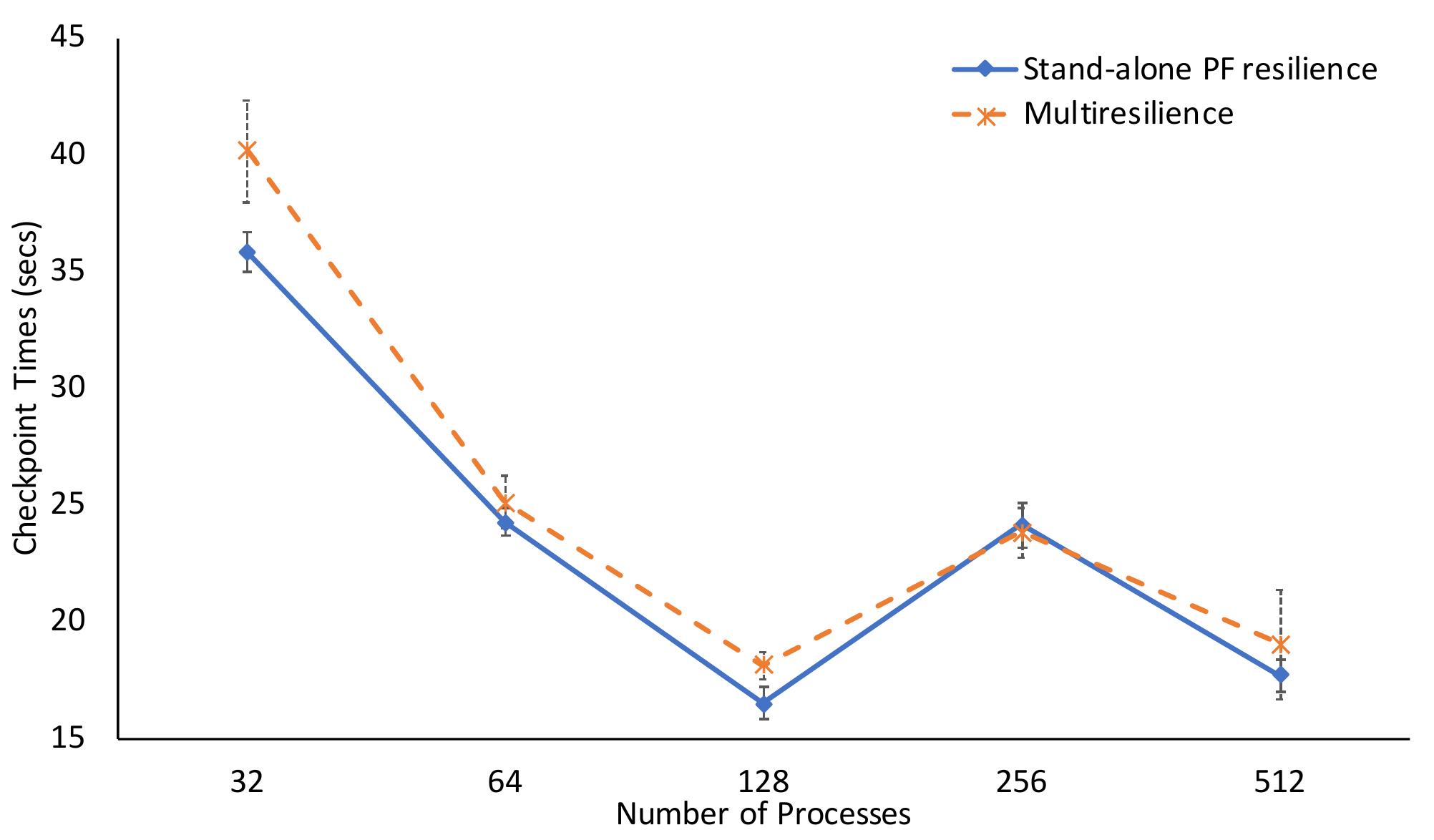}
\caption{Comparison of average checkpoint times between PF resilience and multiresilience solutions.}
\label{Fig:HESEbackups}
\end{figure}

\subsection{Multiresilience Solution}

The performance impact of using the proposed multiresilience solution is quantified in Figure~\ref{Fig:HESEOverall}.
In this figure, we also plot an estimate of the time-to-solution (patterned bars) obtained via results from standalone SE and PF resilience experiments.
The overheads for multiresilience solution tend to decrease with scale in line with the results obtained from PF experiments. 
The observed minute discrepancies between estimated and actual times is due to the interaction between PF and SE resilience patterns which is not captured in stand-alone PF experiments.
Results in Figure~\ref{Fig:HESEbackups} indicate higher average and standard deviation of checkpoint times in multiresilience experiments as compared to stand-alone PF experiments.
The higher checkpoint times are primarily due to extra iterations in multiresilience experiments as a result of SDCs (see Table~\ref{tab:SE-overheads}) causing more checkpoints to be performed.

In our experiments, we are able to achieve decent overall performance estimates for multiresilience solution using stand-alone experiments for two reasons: 1) the cost of performing extra checkpoints of dynamic state due to SDCs is significantly low (see \% dynamic state in checkpoint overhead from Table~\ref{tab:CR-overheads}),
2) the additional time to converge to a solution in presence of SDCs is less than the expected time of process failure, preventing accumulation of additional overheads as a result of process failure. 
This demonstrates that the careful selection of resilience patterns can alleviate some of the complications associated with design of multiresilience solutions for HPC applications.
However, in general, it is important to consider the interactions between different resilience patterns.

\section{Related Work}

There have been numerous proposals for resilience solutions that attempt to solve the challenge at different layers of the system stack. To deal with memory errors, HPC systems use memory modules with error correcting codes (ECC). Algorithm specific schemes, such as algorithm-based fault tolerance (ABFT) apply row and column checksum encoding on dense matrix structures \cite{Huang:1984}, or use the diagonal, banded diagonal, block diagonal structures of sparse problems \cite{Sloan:2012:DSN} for application-level detection and correction of errors. The design of solutions that combine capabilities across different layers of the system stack has also been previously explored, but using ad-hoc methods. For example, using the ABFT technique to protect application data structures permits different ECC mechanisms for different page frames in memory \cite{Li:2013}. To deal with fail-stop and silent errors simultaneously, recent work has proposed combining ABFT methods with system-based checkpointing \cite{Benoit:2015}, in which each computational phase is followed by ABFT verification for SDCs and an in-memory checkpoint. This approach has also been shown to facilitate roll-forward recovery for conjugate gradient solvers \cite{Fasi:2015}.  

Design patterns have been extensively used in software engineering, particularly in the context of object-oriented (OO) programming \cite{Gamma:1995}. Patterns in this context define class interfaces and inheritance hierarchies, and help establish key relationships among the classes. For parallel software design, there have been efforts to codify the various parallel computation and communication structures into a pattern catalog \cite{Mattson:2004}. The Our Pattern Language (OPL) \cite{Mattson:OPL:2009} supports the design and implementation of a parallel algorithm by linking the various parallel design patterns. To the best of our knowledge, this work is the first demonstration of applying resilience design patterns for modeling and implementation of multiresilience solutions.   

\section{Conclusion}
\label{sec:Conclusion}

A pattern-oriented approach to the design and implementation of a multiresilience solution is described in this paper.
We leveraged resilience design patterns to systematically identify and evaluate the appropriate detection and mitigation techniques for two very different fault models: soft errors that cause silent data corruptions and fail-stop failures. 
We demonstrated the development of a multiresilience solution and presented the experimental evaluation for an iterative linear solver application using algorithm-based pattern instances together with patterns realized using ULFM extensions to MPI. 
The broader impact of this work is two fold: it demonstrates a structured model-based approach to identifying alternative patterns for detection, containment and mitigation of specific types of errors, and it facilitates the development of roadmaps for architecting multiresilience solutions by composing patterns from multiple layers of the system stack and iteratively refining the pattern relationships to optimize end-to-end application performance.

\begin{acks}
This material is based upon work supported by the U.S. Department of Energy, Office of Science, Office of Advanced Scientific Computing Research, program manager Lucy Nowell, under contract number DE-AC05-00OR22725.
The authors would like to thank James Elliot from Sandia National Laboratories for his help with the FT-GMRES code.
\end{acks}

\bibliographystyle{ACM-Reference-Format}
\bibliography{references} 

\end{document}